\documentclass[aps,prl,nobibnotes,reprint,longbibliography,floatfix]{revtex4-2}

\usepackage[utf8]{inputenc}
\usepackage[T1]{fontenc}
\usepackage{amsmath,amssymb,latexsym,graphicx}
\usepackage{newtxtext,newtxmath}
\usepackage{bbold,anyfontsize}
\usepackage[dvipsnames]{xcolor}
\usepackage[
  colorlinks=true,
  urlcolor=BlueViolet,
  citecolor=BlueViolet,
  filecolor=BlueViolet,
  linkcolor=BlueViolet
]{hyperref}

\begin{document}

\title{
  Very persistent random walkers reveal transitions in landscape topology
}

\author{Jaron Kent-Dobias}
\email{jaron@ictp-saifr.org}
\affiliation{ICTP South American Institute for Fundamental Research, São Paulo, Brazil}
\affiliation{Instituto de Física Teórica, Universidade Estadual Paulista ``Júlio de Mesquita Filho'', São Paulo, Brazil}

\begin{abstract}
  We study the typical behavior of random walkers on the microcanonical
  configuration space of mean-field disordered systems. Passive walks have an
  ergodicity-breaking transition at precisely the energy density associated
  with the dynamical glass transition, but persistent walks remain ergodic
  at lower energies. In models where the energy landscape is thoroughly
  understood, we show that, in the limit of infinite persistence time, the
  ergodicity-breaking transition coincides with a transition in the
  topology of microcanonical configuration space. We conjecture that this
  correspondence generalizes to other models, and use it to determine
  the topological transition energy in situations where the landscape properties
  are ambiguous.
\end{abstract}

\maketitle

\paragraph{Introduction.}

The notion of an energy landscape and ideas about its geometry and topology
influence our understanding of diverse phenomena including glasses
\cite{Stillinger_1984_Packing}, spin glasses \cite{Castellani_2005_Spin-glass},
proteins \cite{Onuchic_1997_Theory}, evolution \cite{Arnold_2001_The},
ecosystems \cite{Altieri_2021_Properties}, and machine learning \cite{Ballard_2017_Energy, Draxler_2018_Essentially}. As landscapes in high-dimensional
space, one cannot understand their geometry and topology by making a
topographical map. Instead, one must make sense of the low-dimensional shadows they cast,
either through projection along a few important axes \cite{Tenenbaum_2000_A,
Maaten_2008_Visualizing, Moon_2019_Visualizing, Teoh_2020_Visualizing} or by
studying kinds of summary statistics. The most important summary statistic for
understanding complex landscapes is the entropy of their minima and saddle
points, often called complexity \cite{Bray_1980_Metastable, Rieger_1992_The,
Cavagna_1997_An, Ros_2023_High-dimensional}.

The most commonly invoked feature of the complexity is the level at which the
population of minima begins to outnumber the population of saddle points, known
as the threshold energy $E_\text{th}$. This transition is heuristically thought
as a point of landscape
flatness (a geometric property) \cite{Kurchan_1996_Phase} and landscape percolation (a topological property), both important for explaining why the threshold should attract slow asymptotic dynamics \cite{Cugliandolo_1993_Analytical}. However, recent work has called into question
the threshold's monopoly on flatness \cite{BenArous_2019_Geometry, Folena_2020_Rethinking, Kent-Dobias_2023_How, Kent-Dobias_2024_Conditioning, Kent-Dobias_2024_Algorithm-independent} and its significance to dynamics \cite{Folena_2020_Rethinking,
Folena_2021_Gradient, Folena_2023_On} and landscape topology
\cite{Kent-Dobias_2025_On}.

Here, we focus on a question of topology: at what level in the landscape do
typical points in configuration space transition from being smoothly connected
to being isolated? To answer this question, we look at the dynamics of random
walkers confined to a specific energy level. If a random walk from a typical
initial condition can travel arbitrarily far from its starting point, i.e., is
ergodic, we conclude that the level set of the energy, or the microcanonical
configuration space at that energy, is typically connected.

Unfortunately, the converse is not true: random walks are often not ergodic on
a connected configuration space. We will see that for passive walkers, there
is a direct correspondence between microcanonical and canonical dynamics, with
the ergodicity breaking transition for the random walk occurring at precisely
the energy density of the dynamical glass transition. The transition is driven by entropic barriers, not topology: there is nothing
remarkable about the energy landscape at this energy density.

\begin{figure}
  \includegraphics[width=\columnwidth,trim={0 1.6cm 0 0}]{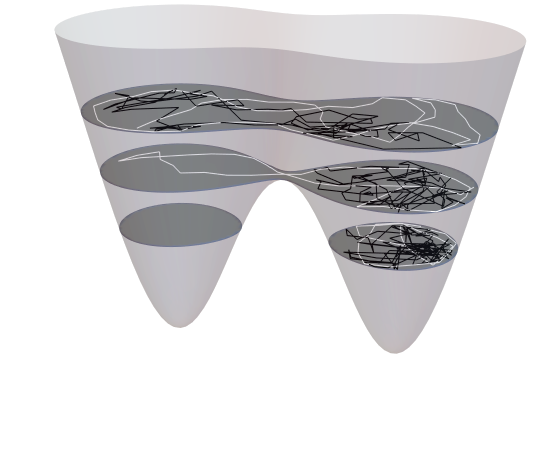}

  \caption{
    The behavior of a passive (black) and persistent (white) random walker on
    level sets of a cartoon landscape. The persistent walker remains ergodic at
    lower energies than the passive one. Increasing persistence pushes the
    ergodic transition of the walk towards a topological transition of the
    landscape associated with the typical connectivity of configurations.
  } \label{fig:walk}
\end{figure}

Fortunately, there are other kinds of random walkers. Persistent or active
random walks, where the walker tends to step in the same direction over a
persistence time $\tau_0$, appear often in biological and biology-inspired settings where the walker expends energy to drive itself forward until its direction is changed either by random noise or by its own volition \cite{Ramaswamy_2017_Active}. Active systems are known to be good at crossing
barriers, entropic or otherwise \cite{Caprini_2019_Active, Woillez_2019_Activated,
Zanovello_2021_Target, Coghi_2025_Accelerated}. Activity is also known to drive the glass transition to
lower temperatures \cite{Berthier_2013_Non-equilibrium, Nandi_2018_A}. In what
follows, we see that persistent walkers remain ergodic at lower energies than
passive ones. Moreover, we argue that, in the limit of infinite persistence
time, the ergodicity breaking transition coincides with the transition in the
landscape topology from typically connected to typically disconnected. A cartoon of this scenario is shown in Fig.~\ref{fig:walk}.

A similar result was recently found in the random Lorentz gas, a simplified
model of the structural glass transition involving a tracer moving freely between
fixed spherical obstacles of some density. When the tracer is passive, there is
a separation between the densities of the dynamical glass transition and the
percolation transition \cite{Biroli_2021_Interplay,
Biroli_2021_Mean-field, Charbonneau_2021_High-dimensional}. On the other hand,
an active tracer sees its ergodicity-breaking transition pushed toward the
percolation transition with increasing activity \cite{Zheng_2025_Not-so-glass-like}. Our results suggest this is the manifestation of a generic phenomenon. Infinitely
persistent activity has also been used to probe properties of the energy
landscape in glasses, both in mean-field and in finite dimensions
\cite{Morse_2021_A, Agoritsas_2021_Mean-field, Mandal_2021_How,
Keta_2023_Intermittent}.

\paragraph{The model.}

We consider the position $\boldsymbol x\in\mathbb R^N$ of a random walker driven by Gaussian noise $\boldsymbol\xi$ with zero mean and with variance
\begin{equation} \label{eq:noise}
  \langle \xi_i(t)\xi_j(s)\rangle=\delta_{ij}\Gamma(t-s)
\end{equation}
for a kernel $\Gamma$ depending only on the time difference. In this paper we typically use an exponentially decaying kernel
\begin{equation} \label{eq:noise.kernel}
  \Gamma(\tau)=\frac1{\tau_0}e^{-|\tau|/\tau_0}
\end{equation}
with characteristic persistence time $\tau_0$. In the limit of zero $\tau_0$,
$\Gamma$ approaches a Dirac $\delta$ function and the noise is Markovian, and we say the walker is passive. The
walker moves freely besides two constraints: it is confined to the sphere
$\|\boldsymbol x\|^2=N$ due to the model, and it is confined to the constant-energy level set $H(\boldsymbol
x)=EN$. Such a random walk can be described by the Langevin equation
\begin{equation} \label{eq:eom}
  \frac{\partial\boldsymbol x(t)}{\partial t}=\boldsymbol\xi(t)-\mu(t)\boldsymbol x(t)-\beta(t)\boldsymbol\nabla H\big(\boldsymbol x(t)\big),
\end{equation}
where $\mu$ and $\beta$ are time-dependent parameters that adjust the magnitude
of forces perpendicular to the constraint manifolds in order to prevent the
walker from leaving them. The dynamics of our walker are equivalent to that of
an Ornstein--Uhlenbeck particle traversing the configuration space
\cite{Martin_2021_Statistical}. Ignoring the spherical constraint, this
framework is general: a system of $P$ active Brownian particles in $d$
dimensions is captured by configurations $\boldsymbol x$ in $N=Pd$ dimensions
concatenating the positions of each particle.

We take our model to be a spherical spin glass, a family of mean-field models
of glassy behavior whose properties are closely related to the random first order
and mode-coupling theories of the glass transition
\cite{Kirkpatrick_1987_p-spin-interaction, Kirkpatrick_1987_Connections,
Kirkpatrick_1987_Stable, Kirkpatrick_2015_Colloquium}.
In equilibrium, they undergo a dynamical transition at that breaks ergodicity without passing an equilibrium phase transition.
The Hamiltonian $H$ is a random
polynomial of the components of $\boldsymbol x$ with independent centered
Gaussian coefficients, or
\begin{equation}
  H(\boldsymbol x)
  =\sum_{p=0}^\infty\frac1{p!}\sqrt{\frac{f^{(p)}(0)}{N^{p-1}}}
  \sum_{i_1,\cdots,i_p}^NJ^{(p)}_{i_1,\cdots,i_p}x_{i_1}\cdots x_{i_p},
\end{equation}
where the $J$ are centered Gaussian with $\overline{J^2}=1$ \cite{Crisanti_1992_The, Crisanti_1993_The}.
The composition of the random polynomial is compactly encoded in a function $f$ whose series coefficient at order $p$ gives the relative strength of the degree-$p$ contribution to $H$. The function $f$ also gives the covariance between $H$ evaluated at two different points in space, with
\begin{equation}
  \overline{H(\boldsymbol x)H(\boldsymbol x')}
  =Nf\left(\frac{\boldsymbol x\cdot\boldsymbol x'}N\right).
\end{equation}
The pure $p$-spin models have $H$ a homogeneous polynomial of degree $p$ and $f(q)=\frac12q^p$,
while all other models are referred to as mixed. In this paper, we refer to equally-appointed mixed models with $f(q)=\frac14(q^p+q^s)$ as $p+s$-spin.

\begin{figure}
  \includegraphics[width=\columnwidth]{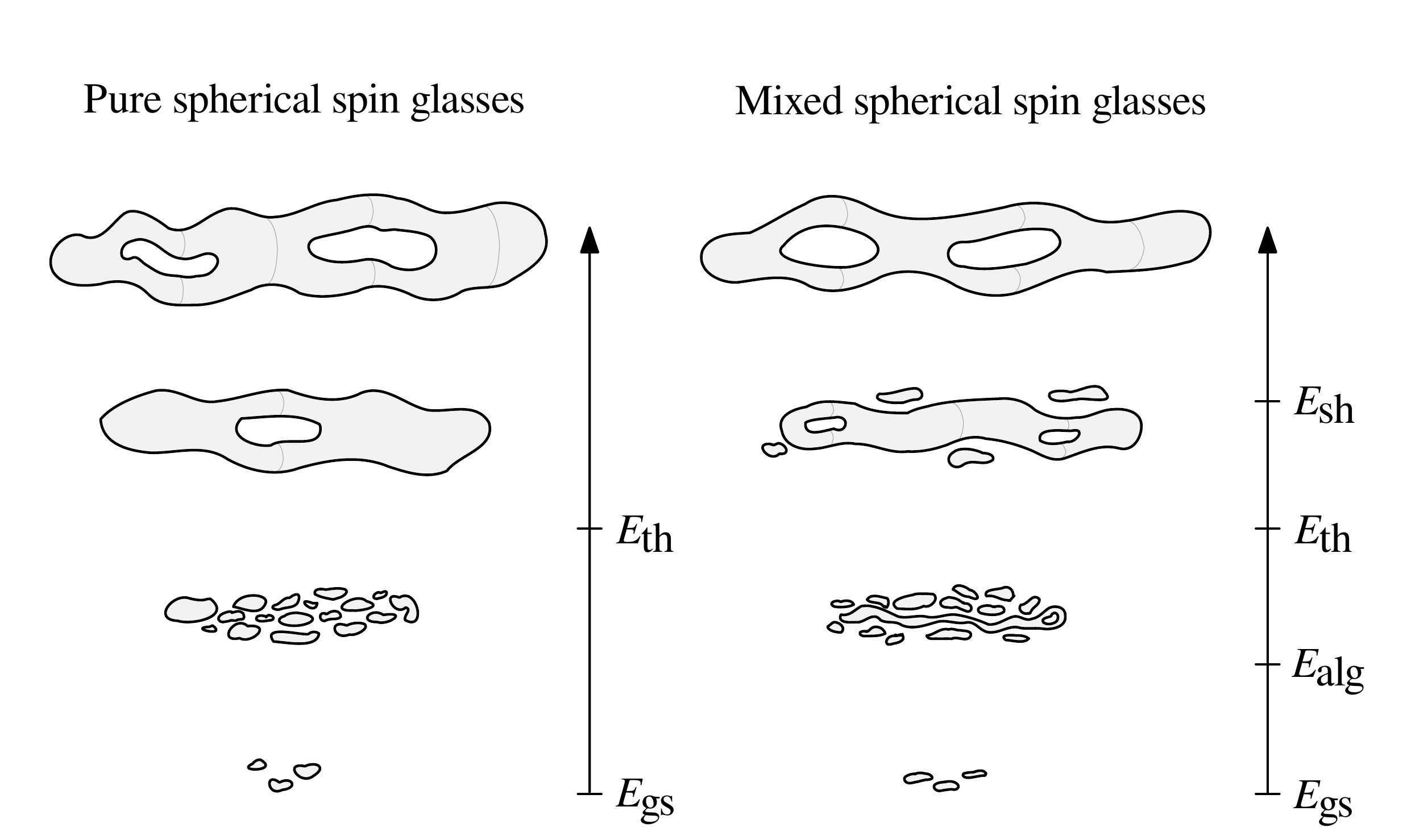}
  \caption{
    Schematic depiction of the energy level set topology of spherical spin
    glasses as energy density is varied. The threshold $E_\text{th}$ is defined
    by the energy at which minima begin to outnumber saddle points. In pure
    models the level set is connected above $E_\text{th}$ while no connected
    component exists below $E_\text{th}$. In mixed models the breaking up of
    the level set occurs over a range of energies. $E_\text{alg}$ is the lowest
    energy at which a large connected component of the level set exists \cite{Gamarnik_2021-10_The}, while
    $E_\text{sh}$ marks where the Euler characteristic of the level set changes
    sign \cite{Kent-Dobias_2025_On}. In pure models, $E_\text{th}=E_\text{alg}=E_\text{sh}$. An aim of
    this manuscript is to determine the lowest energy at which typical parts of
    the level set belong to a large connected component.
  } \label{fig:levels}
\end{figure}

The complexity of these models has been extensively studied \cite{Rieger_1992_The, Crisanti_1995_Thouless-Anderson-Palmer, Cavagna_1997_An, Cavagna_1998_Stationary, Crisanti_2006_Spherical, Auffinger_2012_Random, Auffinger_2013_Complexity, BenArous_2019_Geometry}, and so long as there is not replica symmetry breaking among stationary points \cite{Kent-Dobias_2023_How, Kent-Dobias_2023_When} the threshold energy below which minima outnumber saddle points is
\begin{equation}
  E_\text{th}=-\frac{f'(1)^2+f(1)\big(f''(1)-f'(1)\big)}{f'(1)\sqrt{f''(1)}}.
\end{equation}
In the pure spherical spin glasses the threshold energy is unambiguously
significant in the landscape geometry and topology because it is a sharp
boundary between populations of minima and saddle points, and this is reflected in its importance to out-of-equilibrium dynamics \cite{Cugliandolo_1993_Analytical, Kurchan_1996_Phase}. In mixed models the
boundary is not sharp, with exponentially many minima found above the threshold
and exponentially many saddle points found below it. In this setting, the
topological and geometric significance of the threshold is not obvious. Sketches of the topology of the energy level sets of pure and mixed models are shown in Fig.~\ref{fig:levels}. We find that the ergodicity-breaking transition of
infinitely-persistent random walkers coincides with the threshold in certain
mixed models, conditionally validating its topological significance.

\paragraph{The dynamical equations.}

We seek asymptotic solutions to the correlation and response functions
resulting from the Langevin equation \eqref{eq:eom} averaged over noise $\pmb\xi$ and disorder $J$ in the limit of large $N$.
If a deterministic source $\boldsymbol h(t)$ is added to the equation of
motion, these dynamic order parameters are
\begin{align}
  C(t,s)=\frac1N\overline{\langle\boldsymbol x(t)\cdot\boldsymbol x(s)\rangle}
  &&
  R(t,s)=\frac1N\sum_{i=1}^N\overline{\left\langle\frac{\delta x_i(t)}{\delta h_i(s)}\right\rangle}\bigg|_{\boldsymbol h=0},
\end{align}
respectively. We seek time-translation invariant solutions where the order parameters take the form $C(t+\tau,t)=C(\tau)$, $R(t+\tau,t)=R(\tau)$, $\mu(t)=\mu$, and $\beta(t)=\beta$ independent of $t$.
The random walk will be judged ergodic if the correlation function $C$ decays to zero at large time difference $\tau$.
Employing standard methods detailed in the end matter \cite{Cugliandolo_2002_Dynamics}, the typical behavior of the correlation and
response functions averaged over realizations of the noise and of the disorder
obey integro-differential equations of the form
\begin{widetext}
  \begin{align}
    \label{eq:diff.C}
    \left(\frac\partial{\partial\tau}+\mu\right)C(\tau)
    &=2\int d\sigma\,\Gamma(\tau-\sigma)R(-\sigma)
    +\beta^2\int d\sigma\,R(\tau-\sigma)f''\big(C(\tau-\sigma)\big)C(\sigma)
    +\beta^2\int d\sigma\,f'\big(C(\tau-\sigma)\big)R(-\sigma) \\
    \label{eq:diff.R}
    \left(\frac\partial{\partial\tau}+\mu\right)R(\tau)
    &=\delta(\tau)+\beta^2\int d\sigma\,R(\tau-\sigma)f''\big(C(\tau-\sigma)\big)R(\sigma),
  \end{align}
\end{widetext}
along with equations
\begin{equation} \label{eq:constraint.eqs}
  E=-\beta\int d\tau\,f'\big(C(\tau)\big)R(\tau)
  \hspace{3em}
  1=C(0)
\end{equation}
fixing the values of $\beta$ and $\mu$. Together, these equations imply a
generalized fluctuation--dissipation relation of the form
\begin{equation} \label{eq:fdt}
  R(\tau)=-\Theta(\tau)\int d\sigma\,\Gamma^{-1}(\tau-\sigma)C'(\sigma),
\end{equation}
where $\Theta$ is the Heaviside function and $\Gamma^{-1}$ is the inverse of $\Gamma$ under convolution. For the noise
kernel \eqref{eq:noise.kernel} with persistence time $\tau_0$,
$\Gamma^{-1}(\tau)=\delta(\tau)-\tau_0^2\delta''(\tau)$, which gives
$R(\tau)=-\Theta(\tau)[C'(\tau)-\tau_0^2C'''(\tau)]$. This latter relation resembles ones previously derived for active Ornstein--Uhlenbeck particles \cite{Martin_2021_Statistical, Caprini_2021_Fluctuation-Dissipation}. The generic relation \eqref{eq:fdt} should hold in general for processes driven by noise \eqref{eq:noise} under deterministic forces derived from a potential.

\paragraph{Exact behavior.}

When $\tau_0=0$ and the random walk is passive, these dynamical equations correspond to
those for the canonical equilibrium dynamics of the model at the temperature $T$ whose
average energy density is $E$, but with time rescaled by a constant factor.
This implies the equivalence of the microcanonical and canonical dynamics.
Therefore, the ergodicity-breaking transition of the walker occurs at the
energy density corresponding to the dynamical glass transition temperature $T_\text
d$ where
\begin{equation}
  T_\text d=\sqrt{\frac{1-q_\text d}{q_\text d}f'(q_\text d)}
  \qquad
  1-q_\text d=\frac{f'(q_\text d)}{q_\text df''(q_\text d)},
\end{equation}
which in these models implies $E_\text d=-f(1)/T_\text d$.

\begin{figure}
  \includegraphics{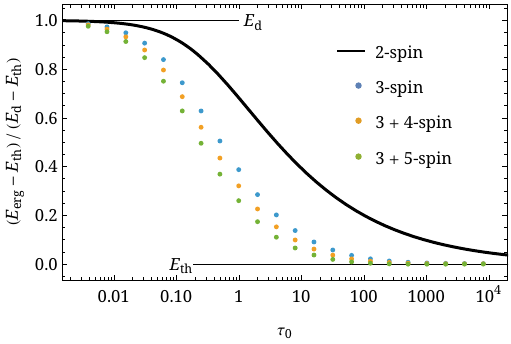}

  \caption{
    The ergodicity-breaking transition energy as a function of persistence time
    $\tau_0$ of an active random walker on the microcanonical configuration
    space of several spherical spin glasses. The solid black line is the
    explicit solution \eqref{eq:ed.2-spin} for the pure 2-spin model, while the
    points show numeric estimates for several other models.
  } \label{fig:crossover}
\end{figure}

The ergodicity-breaking transition for $\tau_0=0$ does not correspond to any
qualitative change in the microcanonical configuration space, but rather is a
result of entropic barriers. We can clearly show this in the situation where
the equations are exactly solvable, for the pure 2-spin model with
$f(q)=\frac12q^2$. Here, the energy landscape is not complex: it has two
symmetrically related minima with energy density $E=-1$, and its microcanonical
configuration space spans all overlaps and has the same topology for all
energy densities $-1<E<0$. Nevertheless, the dynamical glass transition energy with
$\tau_0=0$ is $E_\text d=-\frac12$, an unremarkable level in the landscape.

A persistent walker maintains ergodicity to much lower energy densities. In the solvable 2-spin model, the energy density of the ergodicity-breaking transition is given as a function of $\tau_0$ by
\begin{equation} \label{eq:ed.2-spin}
  E_\text{erg}(\tau_0)=-1+\frac1{\sqrt{3\tau_0}}\sinh\left[\frac13\sinh^{-1}\left(\frac32\sqrt{3\tau_0}\right)\right],
\end{equation}
which interpolates between $E_\text{erg}(0)=E_\text d=-\frac12$ and $E_\text{erg}(\infty)=E_\text{th}=-1$ and is plotted in Fig.~\ref{fig:crossover}.
Therefore, in the pure 2-spin model an infinitely-persistent walker preserves
ergodicity down to the energy density at which a topological change in the
landscape makes ergodicity impossible. In what follows, we argue that this
property continues to hold in other models.

\paragraph{Numeric solutions.}

These equations lack an important feature usually exploited to numerically
solve them: \eqref{eq:diff.C} at a given time depends on the value of $C$ for
all times previous, \emph{including all negative times}. This scenario arises whenever detailed balance is violated \cite{Cugliandolo_1997_Glassy, Berthier_2000_A, Berthier_2013_Non-equilibrium}. Hence one cannot solve the equations by
beginning at $\tau=0$ and stepping forward as is often done. In order to
produce numeric solutions, we solve the equations by iteration starting from
the exact solution for $C$ and $R$ when $E=0$. Details of this procedure can be found in the end matter.

We estimate the energy of the ergodicity-breaking transition for a given
persistence time $\tau_0$ by numerically solving the dynamical equations for
successively smaller $E$ until the Fourier transform of the correlation
function $C$ develops a singularity at $\hat C(\omega=0)$, which signals the
presence of a plateau in the correlation function at nonzero overlap $q$ with
the initial condition. Estimates of these transition energies assuming the divergence of $\hat C(0)$ is a power law like in equilibrium \cite{Leutheusser_1984_Dynamical, Bengtzelius_1984_Dynamics, Caltagirone_2012_Critical} are plotted in
Fig.~\ref{fig:crossover} for pure and mixed models. In both, increasing
$\tau_0$ smoothly interpolates between the dynamical glass transition energy
and the threshold energy where minima begin to outnumber saddle points.

\begin{figure}
  \includegraphics{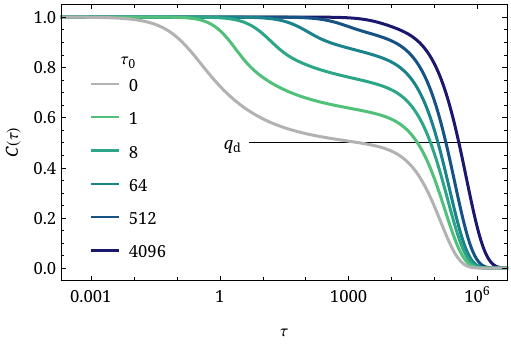}

  \caption{
    The correlation function for the pure
    3-spin spherical spin glass at several values of the persistence time
    $\tau_0$. The plateau overlap $q_\text d$ of the equilibrium dynamical transition of $\tau_0=0$ is marked. As $\tau_0$ increases, the plateau of the ergodicity-breaking
    transition moves to higher overlaps. The energies associated in order of
    increasing $\tau_0$ are $-0.8157$, $-1.0230$, $-1.1076$, $-1.1421$,
    $-1.1517$, and $-1.1538$, whereas $E_\text d\simeq-0.8165$ and
    $E_\text{th}\simeq-1.1547$
  } \label{fig:q}
\end{figure}

As the persistence of the walker increases, not only the energy of the
transition changes, but also the overlap $q=C(\infty)$ associated with the loss of
ergodicity grows towards $q=1$. This is reflected in Fig.~\ref{fig:q}, which
plots $C$ near the transition energy for a variety of persistence times
$\tau_0$. Each correlation function shows a characteristic bump in the decay,
and the increasing value of $C$ at the bump indicates the growing $q$. In the
limit of infinite persistence, the asymptotic overlap at the transition appears
to grow to $q=1$. The ratio $\mu/\beta$ at the transition approaches
$\sqrt{4f''(1)}$ at infinite persistence time, something perhaps related to marginality.

\begin{figure}
  \raggedleft
  \includegraphics{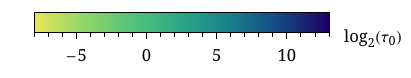}
  \includegraphics{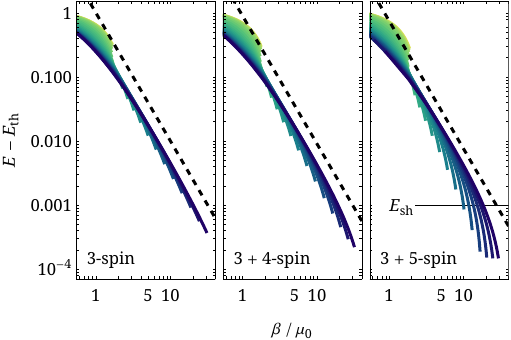}

  \caption{
    Energy difference above the threshold energy as a function of $\beta\mu_0$,
    where $\mu_0$ is the value of $\mu$ associated with a free persistent
    random walk given by \eqref{eq:μ₀}. Each line shows the behavior at fixed
    persistence time for $\tau_0=2^{-8},2^{-7},\ldots,2^{13}$, with larger $\tau_0$ corresponding to lower endpoints. The dashed black
    line is an inverse-square power law $\beta^{-2}$. In the $3+5$-spin model, a horizontal line shows the location of $E_\text{sh}$ defined in Ref.~\cite{Kent-Dobias_2025_On} where another aspect of the landscape topology undergoes a transition.
  } \label{fig:approach}
\end{figure}

We cannot precisely measure the asymptotic energy by fitting $\hat C(0)$ as a
function of $E$ because the power law associated with its divergence is not
known in general. However, in the exact solution for the pure 2-spin model at infinite persistence time, the energy is related to the parameter $\beta$ by
$E=E_\text{erg}+\frac12\beta^{-2}+O(\beta^{-4})$ for $\beta\to\infty$. This
scaling behavior appears to also hold in other models, which allows us to
precisely identify the ergodic transition at infinite persistence time. This increasingly power-law approach of the
energy to the threshold with increasing $\tau_0$ is shown in
Fig.~\ref{fig:approach} for the three nontrivial models studied here. Other
prospective transition energies, like that associated with a topological invariant introduced in
Ref.~\cite{Kent-Dobias_2025_On}, are clearly ruled out. In
all three cases, our data suggests that the threshold energy $E_\text{th}$ is
the ergodic transition energy in the infinitely-persistent limit.

In the pure spherical models the dynamic, geometric, and topological
significance of the threshold energy is not ambiguous. The fact that at
infinite persistence time the ergodic transition reaches this topologically
significant point leads us to conjecture that this behavior is generic: that
the energy level of the ergodicity-breaking transition for an
infinitely-persistent walker is topologically significant, in the sense that
under it typical points in configuration space do not belong to the same connected
component. Based on this conjecture and the numeric data discussed above, we therefore conclude that this topological transition takes place at $E_\text{th}$ also in the mixed spherical models examined here.

\paragraph{Conclusions.}

We have developed a dynamical mean field theory of persistent random walks on
the microcanonical configuration space of models with complex landscapes. Based
on an exact solution and on approximate numeric solutions for the
well-understood pure $p$-spin spherical models, we conjecture that an
infinitely persistent walker remains ergodic until a topological transition in
the configuration space is passed. By further examining numeric solutions on
the less-understood mixed spherical models, we argue that the
ergodicity-breaking transition at infinite persistence time (and therefore the
topological transition) occurs at the threshold energy, where minima first
outnumber saddle points.

Given the controversy surrounding the threshold energy described in the
introduction, it may be surprising that we find it is the level at which the
microcanonical configuration space becomes typically disconnected. However,
very recent work suggests that prior conclusions questioning its dynamic
significance may have been based on transient behavior, at least for the models
we describe here \cite{Lang_2025_Numerical, Lang_2025_Personal}. But
$E_\text{th}$ cannot hold its dynamic significance in \emph{all} models,
because in some cases it can fall beneath the energy $E_\text{alg}$ proven to
bound the performance of polynomial-time algorithms, including all physical
dynamics \cite{ElAlaoui_2020_Algorithmic, ElAlaoui_2021_Optimization}. Since
$E_\text{alg}$ also is the lowest energy at which \emph{any} connected
component of the level set exists (see Fig.~\ref{fig:levels}), the topological
transition studied here must also differ from $E_\text{th}$ for these models
\cite{Gamarnik_2021-10_The}. The importance of $E_\text{alg}$ results from the difficulty of dynamics to navigate the landscape in the absence of a large connected component; given the elementary greedy nature of gradient descent, we suspect it stalls when a large connected component becomes atypical even when this does not correspond to $E_\text{th}$. Understanding the nature of the energy landscape
and its relationship to out-of-equilibrium dynamics in such models is a
priority for future research. We do not expect the threshold energy usually
holds this topological significance in more structured mean-field models and in
finite-dimensional systems.

The conjectured connection between infinite persistence and topological
transitions in configuration space should be studied in systems outside of mean
field. Already, existing work on the random Lorentz gas is consistent with this
conjecture, though it is also a toy model \cite{Zheng_2025_Not-so-glass-like}.
Making a similar analysis of hard or soft spheres may be restricted by the
current understanding of the topological properties of their configuration
space; detailed analysis currently exists only for systems of relatively few
particles, for which all walks are ergodic \cite{Carlsson_2012_Computational,
Barnett-Jones_2013_Transition, Weeks_2020_Visualizing,
Ericok_2021_Configuration}.
In the absence of direct topological approaches, this method may advance our
understanding of the topology of configurations in particle systems. Numeric
techniques developed for simulating infinitely persistent activity
\cite{Mandal_2021_How} may permit the same analysis in finite-dimensional
systems, like active hard- and soft-sphere glasses
\cite{Keta_2023_Intermittent}. Mean field liquids may admit an analytic approach, extending the existing dynamical
mean-field theory for infinitely persistent active fluids
\cite{Agoritsas_2021_Mean-field} by adding a microcanonical energy constraint.

\paragraph{Acknowledgements.}

The authors thank James P Sethna, Ralph B Robinson, and Bethany Dixon at
Cornell University for providing and facilitating access to computing
resources used in this work, and thank Johannes Lang for useful conversations.
JK-D is supported by FAPESP Young Investigator Grant No.~2024/11114-1. JK-D
also received support from the Simons Foundation Targeted Grant to ICTP-SAIFR.

\paragraph{Data availability.}

The data that support the findings of this article are openly available \cite{Kent-Dobias_2025_Log-Fourier, Kent-Dobias_2026_Data}.

\bibliography{threshold}

\onecolumngrid

\section{End Matter}

\twocolumngrid

\paragraph{Deriving the dynamical equations.}

The procedure for writing the dynamical equations above is not novel \cite{Cugliandolo_2002_Dynamics}, but it is worth following aspects of it carefully to see how the generalized fluctuation--dissipation relation arises.
Solutions to the Langevin equation \eqref{eq:eom} can be sampled using the path integral
\begin{align}
  &Z
  =\int\mathcal D\boldsymbol x\,\mathcal D\mu\,\mathcal D\beta\,
    \delta\big(\boldsymbol\xi_t-\partial_t\boldsymbol x_t-\mu_t\boldsymbol x_t-\beta_t\boldsymbol\nabla H(\boldsymbol x_t)\big) \\
  &\quad\times\delta\big(\tfrac12(N-\|\boldsymbol x_t\|^2)\big)\,
    \delta\big(NE-H(\boldsymbol x_t)\big) \notag \\
  &\times
      \det
      \begin{bmatrix}
        [I(\partial_t+\mu_t)+\beta_t\boldsymbol\nabla^2H(\boldsymbol x_t)]\delta_{ts}
        &\boldsymbol x_t\delta_{ts} & \boldsymbol\nabla H(\boldsymbol x_t)\delta_{ts} \\
        \boldsymbol x_s\delta_{ts} & 0 & 0 \\
        \boldsymbol\nabla H(\boldsymbol x_s)\delta_{ts} & 0 & 0
      \end{bmatrix}.
     \notag
\end{align}
The integrand is converted to an exponential function by writing
the $\delta$ functions in their Fourier representations and by writing the
determinant with Grassmann variables. In this form the noise can be averaged
away. The result is compactly represented with superspace coordinates:
introducing Grassmann indices $\bar\theta$ and $\theta$ and defining
$a=(t,\theta,\bar\theta)$, we write
\begin{align}
  \boldsymbol\phi_a&=\boldsymbol x_t+\bar\theta\boldsymbol\eta_t+\bar{\boldsymbol\eta}_t\theta+\bar\theta\theta\hat{\boldsymbol x}_t \\
  B_a&=\beta_t+\bar\theta\gamma_t+\bar\gamma _t\theta+\bar\theta\theta\hat\beta_t \\
  \Lambda_a&=\mu_t+\bar\theta\vartheta_t+\bar\vartheta_t\theta+\bar\theta\theta\hat\mu_t
\end{align}
for Grassmann fields $\bar{\boldsymbol\eta}$, $\boldsymbol\eta$,
$\bar\gamma$, $\gamma$, $\bar\vartheta$, and $\vartheta$, and auxiliary
real-valued fields $\hat{\boldsymbol x}$, $\hat\beta$, and $\hat\mu$.
The result is
\begin{align}
  \langle Z\rangle
  =\int\mathcal D\boldsymbol\phi\,\mathcal DB\,\mathcal D\Lambda\,
  \exp\bigg\{
    \int da\,\bigg[\frac12\boldsymbol\phi_a^TD^{(2)}_a\boldsymbol\phi_a \qquad \notag\\
      +\frac12\Lambda_a(N-\|\boldsymbol\phi_a\|^2)
      +B_a\big(NE-H(\boldsymbol\phi_a)\big)
    \bigg]
  \bigg\}.
  \label{eq:path.integral}
\end{align}
The differential operator that produces the kinetic part of the action is
defined by
\begin{equation}
  D^{(2)}_a\psi
  =2\Gamma\ast\frac{\partial^2\psi}{\partial\theta\partial\bar\theta}
  +2\theta\frac{\partial^2\psi}{\partial\theta\partial t}
    -\frac{\partial\psi}{\partial t},
\end{equation}
where $\ast$ denotes convolution in the time coordinate. Averaging over the
disorder in $H$ and introducing order parameters
$Q_{ab}=\frac1N\boldsymbol\phi_a\cdot\boldsymbol\phi_b$, we find an effective
action
\begin{align}
  &\mathcal S(Q,B,\Lambda)=
    \int da\,\left[
      \frac12\Lambda_a
      +B_aE
    \right]
    +\frac12\log\det Q
    \\
  &\qquad+\frac12\int da\,db\,\left[
      \delta_{ab}(D^{(2)}_a-\Lambda_a)Q_{ab}
      +B_aB_bf(Q_{ab})
    \right].
    \notag
\end{align}
Differentiating with respect to the order parameters to extremize the action results in the equations
\begin{align}
  \label{eq:eq.Q}
  0&=(D^{(2)}_a-\Lambda_a)Q_{ab}+\int dc\,B_aB_cf'(Q_{ac})Q_{cb}+I_{ab} \\
  \label{eq:eq.ΛB}
  0&=\int db\,B_bf(Q_{ab})+E \hspace{4em}
  0=Q_{aa}-1.
\end{align}
The action of the path integral \eqref{eq:path.integral} is invariant under the action of the set of operators \cite{Kurchan_1992_Supersymmetry}
\begin{align}
  \bar D'_a\psi=\Gamma\ast\frac{\partial\psi}{\partial\theta}+\bar\theta\frac{\partial\psi}{\partial t}
  &&
  D_a\psi=\Gamma\ast\frac{\partial\psi}{\partial\bar\theta}-\theta\frac{\partial\psi}{\partial t} \\
  D'_a\psi=\frac{\partial\psi}{\partial\bar\theta}
  &&
  \bar D_a\psi=\frac{\partial\psi}{\partial\theta}
  \\
  [\bar D_a', D_a']_+\psi=\frac{\partial\psi}{\partial t}
  &&
  [\bar D_a, D_a]_+\psi=-\frac{\partial\psi}{\partial t}.
\end{align}
In particular the kinetic term $D_a^{(2)}=[\bar D_a,D_a]_-$. The symmetry with
respect to these operators and their two-component generalizations $\boldsymbol
D'_{ab}=D'_a+D'_b$, $\bar{\boldsymbol D}'_{ab}=\bar D'_a+\bar D'_b$, and
$[\bar{\boldsymbol D}'_{ab},\boldsymbol D'_{ab}]_+=\frac\partial{\partial
t}+\frac\partial{\partial s}$ imply Ward identities constraining the order
parameters. The resulting Ward identities imply the vanishing of all fermionic
order parameters, $\hat\mu$ and $\hat\beta$, and the order parameter
corresponding to $\hat{\boldsymbol x}\cdot\hat{\boldsymbol x}$. They also imply
time-translation invariance, and the fluctuation--dissipation relation \eqref{eq:fdt} is implied by the Ward identity associated
with $\bar{\boldsymbol D}'$. Expanding the superspace notation of
\eqref{eq:eq.Q} and \eqref{eq:eq.ΛB} and applying these Ward identities results
in \eqref{eq:diff.C}, \eqref{eq:diff.R}, and \eqref{eq:constraint.eqs}, with
$C(\tau)=\frac1N\boldsymbol x_{t+\tau}\cdot\boldsymbol x_{t}$ and
$R(\tau)=\frac1N\boldsymbol x_{t+\tau}\cdot\hat{\boldsymbol x}_t$.

\paragraph{Exact solution for the 2-spin model.}

For the 2-spin model, with $f(q)=\frac12q^2$, the dynamical equations are
exactly solvable. Written in Fourier space, they are
\begin{align}
  (i\omega+\mu)\hat C(\omega)&=2\hat\Gamma(\omega)\hat R(\omega)^\dagger
  +\beta^2\hat C(\omega)\big[\hat R(\omega)+\hat R(\omega)^\dagger\big] \notag \\
  (i\omega+\mu)\hat R(\omega)&=1+\beta^2\hat R(\omega)^2.
\end{align}
This is quadratic in $\hat C$ and $\hat R$ and can be directly solved. The
general solution is not helpful to share here, but the ergodicity breaking transition occurs when $\beta=\frac12\mu$, where $\hat C$ develops a cusp singularity at $\omega=0$ and takes the form
\begin{align}
  &\hat C(\omega)=\frac1{\mu^3\omega^2(1+\tau_0^2\omega^2)} \\
  &\times\left[
    \left(\omega^2+|\omega|\sqrt{4\mu^2+\omega^2}\right)
    \sqrt{2|\omega|\sqrt{4\mu^2+\omega^2}-2\omega^2}
    -4\mu\omega^2
  \right]. \notag
\end{align}
The spherical constraint requires
\begin{equation}
  1=C(0)=\frac1{2\pi}\int d\omega\,\hat C(\omega)=2\frac{\sqrt{1+2\tau_0\mu}-1}{\tau_0\mu^2},
\end{equation}
which implies that
\begin{equation}
  \mu=4\sqrt{\frac1{3\tau_0}}\sinh\left[\frac13\sinh^{-1}\left(\frac32\sqrt{3\tau_0}\right)\right].
\end{equation}
The energy of the transition is then
\begin{equation}
  E_\text d=-\frac\mu2\int d\tau\,C(\tau)R(\tau)
  =-\frac\mu{4\pi}\int d\omega\,\hat C(\omega)\hat R(\omega),
\end{equation}
which when evaluated produces the formula in the text. We can also work directly at $\tau_0=\infty$. Scaling $\omega$ with $\tau_0$, we find
\begin{equation}
  \lim_{\tau_0\to\infty}\hat C(\nu/\tau_0)
  =\frac1{\beta^2}\frac{\mu\left[\mu-\sqrt{\mu^2-4\beta^2}\right]-4\beta^2}{(4\beta^2-\mu^2)(1+\nu^2)}.
\end{equation}
This is the Fourier form for a simple exponential. $C(0)=1$ gives
\begin{equation}
  \mu=\frac{1+2\beta^2}{\sqrt{1+\beta^2}},
\end{equation}
and then for the energy we have
\begin{equation}
  \lim_{\tau_0\to\infty}\hat C(\nu/\tau_0)\hat R(\nu/\tau_0)
  =\frac2{\sqrt{1+\beta^2}}\frac1{1+\nu^2},
\end{equation}
yielding $E=-(1+\beta^{-2})^{-\frac12}$. The dynamic transition occurs at $\beta=\infty$. This gives the scaling $E=-1+\frac12\beta^{-2}+O(\beta^{-4})$ cited in the main text.

\paragraph{Iteration scheme for numeric solutions.}

\begin{figure}
  \includegraphics[width=\columnwidth]{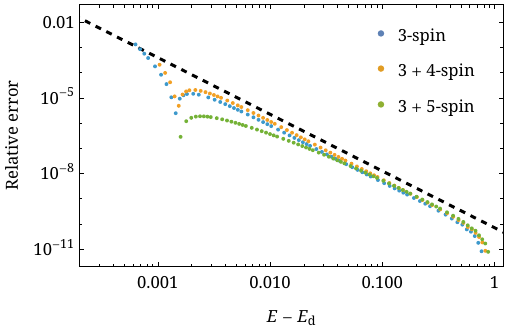}
  \caption{
    Estimation of error in the numeric method using results for $\tau_0=0$,
    where the equilibrium dynamics imply $E=-f(1)\beta$ and the error
    $|\tilde E-E|/|E_\text d-E|$ relative to the distance from the dynamic transition $E_\text d$ can be measured. The relative error grows as $E_\text d$ is approached. The dashed black line shows $(E-E_\text
    d)^{-2.25}$ and is meant as a guide. The smallest differences
    $E-E_\text{th}$ used in this paper are around $2\times 10^{-4}$, where the
    extrapolated error is around 2\%.
  } \label{fig:error}
\end{figure}

To find numeric solutions to the dynamical equations, we use an iterative scheme. We start from the exact solutions for $C$ and $R$ at $E=\beta=0$,
\begin{equation}
  C_0(\tau)=\frac{\mu_0^{-1}e^{-\mu_0\tau}-\tau_0e^{-\tau/\tau_0}}{1-\mu_0^2\tau_0^2}
  \qquad
  R_0(\tau)=\Theta(\tau)e^{-\mu_0\tau},
\end{equation}
where
\begin{equation} \label{eq:μ₀}
  \mu_0=\frac{\sqrt{1+4\tau_0}-1}{2\tau_0}.
\end{equation}
To speed convergence, $\beta$ is fixed and a numeric estimate $\tilde E$ for the energy determined after convergence.
At each step, the self-energies are computed using the current estimates of $C$ and $R$ using
\begin{equation}
  \Sigma_n(\tau)=\beta^2R_n(\tau)f''\big(C_n(\tau)\big)
  \quad
  \mathrm D_n(\tau)=\beta^2f'\big(C_n(\tau)\big).
\end{equation}
Fourier transforming these functions, we then set $C$ and $R$ for the next iteration through their Fourier transforms
\begin{align}
  &\hat C_{n+1}(\omega)=
  \frac{
    \big[2\hat\Gamma(\omega)+\hat{\mathrm D}_n(\omega)\big]\hat R_n(\omega)^\dagger
    +\hat\Sigma_n(\omega)\hat C_n(\omega)
  }
  {i\omega+\mu_{n+1}}
    \notag \\
  &\hat R_{n+1}(\omega)
  =\frac{1+\hat\Sigma_n(\omega)\hat R_n(\omega)}
    {i\omega+\mu_{n+1}},
\end{align}
where $\mu_{n+1}$ is set by the requirement that $C_{n+1}(0)=1$ and is determined by binary search.
The iteration is continued until the integrated root mean square of the dynamical equations is less than some threshold.
We discretize the Fourier transform using logarithmic spacing of $\omega$, which
strongly benefits the precision of the result \cite{Haines_1988_Logarithmic, Lang_2019_Fast}. The method appears to be fundamentally unstable for models containing any 2-spin component, something long known for iterative methods \cite{Cugliandolo_2025_Personal}.

For the data presented here, we represented the correlation and response
functions using 80-bit floating point numbers on an evenly log-spaced grid of
$2^{15}$ points spanning the interval
$(5.9\times10^{-15},1.7\times10^{14})/\mu_0$. The Fourier transforms were made on a domain of $4\times 2^{15}$ points, padded with the time-reflected function and zeroes. After every step, $C$ and $R$ were cropped in the range $[0,1]$ and monotony enforced. The iteration was considered
converged when the \textsc{rms} average over the frequency grid of the difference
between the right- and left-hand sides of equations \eqref{eq:diff.C} and
\eqref{eq:diff.R} fell below $10^{-13}$.
Fourier transforms were carried out using the \textsc{fftw} library
\cite{Frigo_2005_The}. The code used is publicly available
\cite{Kent-Dobias_2025_Log-Fourier}. An estimate of the error due to the method
with these parameters is shown in Fig.~\ref{fig:error}.

\end{document}